\documentclass[useAMS,usenatbib,usegraphicx,usedcolumn]{mn2e}

\title[$\varepsilon$-mechanism instability of metal-poor low-mass stars]{Dipole low-order g mode instability of metal-poor low-mass main-sequence stars due to the $\varepsilon$-mechanism}
\author[T. Sonoi and H. Shibahashi]{Takafumi Sonoi$^{1}$\thanks{E-mail:
sonoi@astron.s.u-tokyo.ac.jp} and Hiromoto Shibahashi$^{1}$\thanks{E-mail:
shibahashi@astron.s.u-tokyo.ac.jp} \\
$^{1}$Department of Astronomy, Graduate School of Science, The University of Tokyo, Hongo 7-3-1, Bunkyo-ku, Tokyo, Japan 113-0033}
\begin{document}

\date{}

\pagerange{\pageref{firstpage}--\pageref{lastpage}} \pubyear{2002}

\maketitle

\label{firstpage}

\begin{abstract}
We analyzed vibrational stability of metal-poor low-mass main-sequence stars due to the $\varepsilon$-mechanism. Since outer convection zones of the metal-poor stars is limited only to the very outer layers, the uncertainty in treatment of convection does not affect the result significantly. We found that the dipole g$_1$- and g$_2$-modes certainly become unstable due to the $\varepsilon$-mechanism for $Z\la 6\times 10^{-4}$. Besides that, we found that as the metallicity decreases the mass range of the $\varepsilon$-mechanism instability extends toward higher mass.  
\end{abstract}

\begin{keywords}
stars: abundances -- stars: low-mass -- stars: oscillations ({\it including pulsations}) -- stars: Population II
\end{keywords}

\section{Introduction}
\label{sec:introduction}
Structure and evolution of stars born in metal-poor environments or when heavy elements were siginificantly deficient in the Universe are considerably different than for stars born much later with higher heavy-element abundances. This is mainly because opacities and nuclear reactions are highly dependent on metallicity. Opacity decreases with metallicity because of lack of absorption by heavy elements. Hence it makes luminosity of the star higher. To maintain energy equilibrium in this situation, the stars with low metallicity need to be compact compared with the metal-rich stars. As a result, the main-sequence of metal-poor stars on the HR diagram moves toward the bluer and higher luminosity side. With this, the convective envelope of stars with $\simeq 1M_{\sun}$ becomes thiner and limited to the very outer layer close to the surface \citep[][hereafter Paper I]{Sonoi2011}. 

The decrease in metallicity also makes CNO-cycle energy generation less efficient. In the Population I case, the pp-chain is dominant for $\la 1.2M_{\sun}$, while the CNO-cycle is influential for the more massive stars. But as the metallicity and the CNO abundance decreases, the CNO-cycle becomes replaced by the pp-chain for more massive stars. In the Population III case, the nuclear energy source is only the pp-chain at the ZAMS stage for stars with $\la 13M_{\sun}$ \citep[][hereafter Paper II]{Sonoi2012}.  

In the case of the pp-chain burning, we should note the vibrational instability of low-degree low-order g modes due to the $\varepsilon$-mechanism. Such instability and the resultant material mixing was once proposed as a possible solution to the solar neutrino problem \citep{Dilke1972}, and the detailed numerical calculations of linear stability analyses demonstrated that such instability is likely to occur in a certain early evolutionary stage of the Sun and the solar-like stars \citep{Chris1974, Boury1975, Shibahashi1975, Noels1976}. The presence of a convective envelope, which occupies the outer 20--30\% of the stellar radius, however, has made it hard to reach a definite conclusion on the vibrational stability because of uncertainty in treatment of the convective envelope. 

However, the situation is different in metal-poor stars. As described above, with decreased metallicity, convective envelopes of the solar-like stars become thin enough that the uncertainty in treatment of convection does not significantly affect results of the vibrational stability analyses.   

Indeed, the present authors showed that metal-free Population III stars are vibrationally unstable against dipole low-order g modes due to the $\varepsilon$-mechanism of the pp-chain (Papers I and II). To extend our analyses to stars with $Z\ne 0$ but still having only very thin convective envelope, we examined the stability of stars with low metallicity by a non-adiabatic analysis, and determined the upper limit metallicity for which this instability appears without the uncertainty.

\section{Evolutionary equilibrium models}
\label{sec:evolutionary}

\begin{figure}
  \includegraphics[width=84mm]{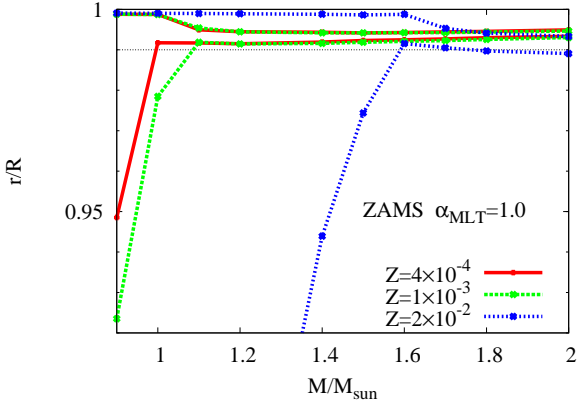}
  \caption{Top and bottom boundaries of an outer convection zone for ZAMS models with $Z=4\times 10^{-4}$, $1\times 10^{-3}$ and $2\times 10^{-2}$. The initial hydrogen abundance is set to be $X_0=0.75$. ({\it A color version of this figure is available in the online journal}).}
  \label{fig:1}
\end{figure}

\begin{figure}
  \includegraphics[width=84mm]{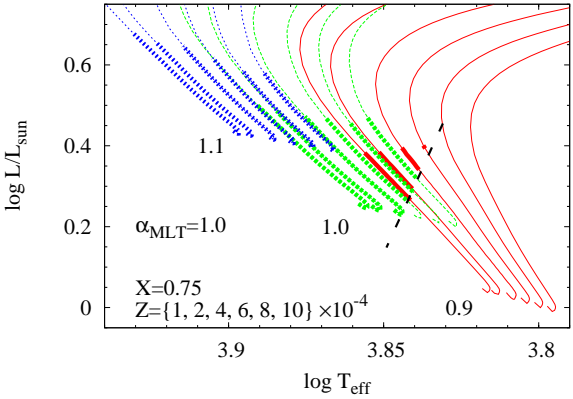}
  \caption{Evolutionary tracks for $0.9$--$1.1M_{\sun}$ with $X_0=0.75$,  $Z=1, 2, 4, 6, 8\times 10^{-4}$ and $1\times 10^{-3}$. For evolutionary models on the thick parts of the tracks, the convective envelope is limited to the outer layer above $r/R=0.99$. The thick dashed line means the red edge of the region where such models are located ({\it A color version of this figure is available in the online journal}).}
  \label{fig:2}
\end{figure}

\begin{figure}
  \begin{center}
    \includegraphics[width=84mm]{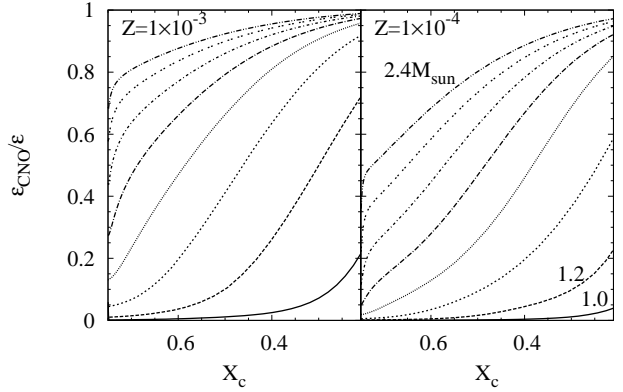}
  \end{center}
    \caption{The fraction of nuclear energy by the CNO-cycle to the whole nuclear energy at the stellar center for $1.0$--$2.4M_{\sun}$ with a 0.2$M_{\sun}$ step. The abscissa is the central hydrogen abundance $X_c$, which is an indicator of stellar evolution. This range is set to be from $X_c=0.75$ to 0.20.}
    \label{fig:3}
\end{figure}

We adopted the same code as in Paper II, MESA \citep{Paxton2011}, to calculate stellar evolution. Evolutionary models were constructed by calculating from the pre--main-sequence Hayashi phase with the initial hydrogen abundance, $X_0=0.75$, and the metallicities, $Z=1, 2, 4, 6, 8\times 10^{-4}$ and $1\times 10^{-3}$. The mixing length parameter was set to be $\alpha_{\rm MLT}=1.0$.  

Fig. \ref{fig:1} shows top and bottom boundaries of an outer convection zone for ZAMS models with different metallicity.
As described in Sec. \ref{sec:introduction}, with decreasing metallicity, outer convection zones for $\simeq 1M_{\sun}$ stars become thinner, limited to outer layers above $r/R\simeq 0.99$, and as thin as or thinner than of population I stars with 1.6--2.0$M_{\sun}$, which correspond to the blue edge of the $\delta$ Scuti instability strip. The vibrational stability of those stars has been well-determined without uncertainty in treatment of convection \citep[e.g.][]{Stellingwerf1979, Tsvetkov1983, Pamyatnykh2000}.
In this study, we restricted our stability analysis to evolutionary models having very thin convective envelopes which are confined to the outer layers above $r/R=0.99$. Fig. \ref{fig:2} shows evolutionary tracks for $0.9$--$1.1M_{\sun}$, and the thick parts of the tracks correspond to such models. The thick dashed line means the red edge of the region where such models are located.   

On the other hand, properties of nuclear reactions also change with metallicity. Fig. \ref{fig:3} compares contribution of the CNO-cycle to the whole nuclear energy generation at the stellar center between $Z=1\times 10^{-3}$ and $1\times 10^{-4}$. The contribution of the CNO-cycle decreases with the metallicity for the same mass star. Energy generated by the pp-chain occupies more than 80\% for $\la 2.2M_{\sun}$ with $Z=1\times 10^{-4}$ at the ZAMS stage, while for $\la 1.6M_{\sun}$ with $Z=1\times 10^{-3}$.

\section{Linear fully nonadiabatic stability analysis} \label{sec:nonadiabatic}
We performed a linear fully nonadiabatic analysis for the above equilibrium models by using a Henyey-type code developed in Paper II. 
By following \citet{Unno1989}, we linearized the equations of continuity, motion, energy conservation and radiative diffusion, and the Poisson's equation, while any perturbation terms are expressed in terms of a combination of a spatial function and a time varying function. The latter is expressed by $\exp (i\sigma t)$, where $\sigma$ denotes the eigenfrequency. The former, the spatial part, is decomposed into a spherical harmonic function, which is a function of the colatitude and the azimuthal angle, and a radial function. Equations governing the radial functions lead to a set of sixth order differential equation, of which coefficients are complex, including terms of frequency $\sigma$. Together with proper boundary conditions, this set of the equations forms a complex eigenvalue problem with an eigenvalue $\sigma$. The real part of $\sigma$, $\sigma_{\mathrm{R}}$, represents the oscillation frequency, and the imaginary part of $\sigma$, $\sigma_{\mathrm{I}}$, gives the growth rate or the damping rate, depending on its sign. We adopted the ``frozen-in convection'' approximation, that is, simply neglected convective flux perturbation. This approximation is acceptable to this study since outer convection zones of the chosen models are very thin and the result of the stability analysis may not depend significantly on treatment of convection.

For a stability analysis relevant to the $\varepsilon$-mechanism, we should adopt temperature and density dependences of nuclear reactions through the perturabation, $\varepsilon_T\equiv(\partial\ln\varepsilon/\partial\ln T)_\rho$ and $\varepsilon_\rho\equiv(\partial\ln\varepsilon/\partial\ln\rho)_T$, which are different from those for evolutionary time-scale. For example, as for the pp-I branch in the pp-chain, the tempearture dependence of energy generation all through the branch in equilibrium is governed by the slowest reaction, $^1$H($^1$H,$e^+\nu_e$)$^2$H, and $\varepsilon_T\simeq 4$ at $\log T=7$. But we should adopt the effective temperature dependence of nuclear reaction through the perturbation, which is mainly governed by $^3$He($^3$He,$2^1$H)$^4$He, and $\varepsilon_T\simeq 11$ \citep{Dilke1972, Boury1973, Unno1975, Unno1989}. We evaluate such effective dependences of the pp-chain and the CN-cycle through the perturbation, separately, and then average them with their contribution to the total nuclear energy generation to get the net values. More details are described in Paper II. 

\section{Results}
\subsection{Variation of stability with stellar evolution and stellar mass dependence}
\label{sec:variation}

\begin{figure}
  \begin{center}
    \includegraphics[width=84mm]{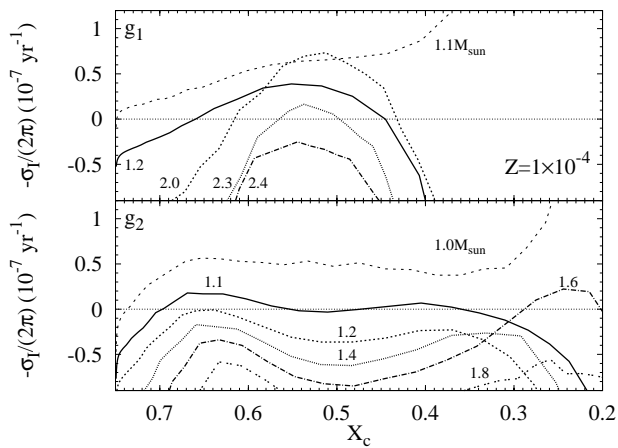}
  \end{center}
    \caption{Variation of growth rates of the dipole($l=1$) g$_1$- (top) and g$_2$-modes (bottom) for different mass stars with $Z=1 \times 10^{-4}$. The abscissa is same as figure \ref{fig:3}.}
    \label{fig:4}
\end{figure}  

In this section, we discuss variation of stability with stellar evolution and stellar mass dependence of the stability while showing results in the $Z=1\times 10^{-4}$ case. Fig. \ref{fig:4} shows variation of growth rates of the dipole ($l=1$) g$_1$- and g$_2$-modes for stars with $Z=1\times 10^{-4}$. The positive value of $-\sigma_{\rm I}/(2\pi)$ means instability. Except for the lower mass stars, the g modes are stable at the ZAMS stage, become unstable in the middle stage of the core hydrogen burning, and eventually become stable again. As well as the Population III case introduced in Paper I and II, the g$_1$- and g$_2$-modes become most unstable or least stable at $X_c=0.5$--$0.6$ and at $X_c=0.6$--$0.7$, respectively. This delicate change of stability is caused by the variation in amplitude distribution with stellar evolution. 
 
In the case of the pp-chain burning, at the stellar centre $^3$He is consumed by $^3$He($^3$He,2$^1$H)$^4$He or by $^3$He($^4$He,$\gamma$)$^7$Be immediately after it is generated. However, since the $^3$He reactions are highly sensitive to temperature, it does not efficiently occur in the outer part of the nuclear burning core. As a consequence, $^3$He accumulates in an off-centered shell. Hence the most favorable situation for the vibrational instability is that the temperature perturbation has a large amplitude in such an off-centered $^3$He shell. 

\begin{figure}
  \begin{center}
    \includegraphics[width=84mm]{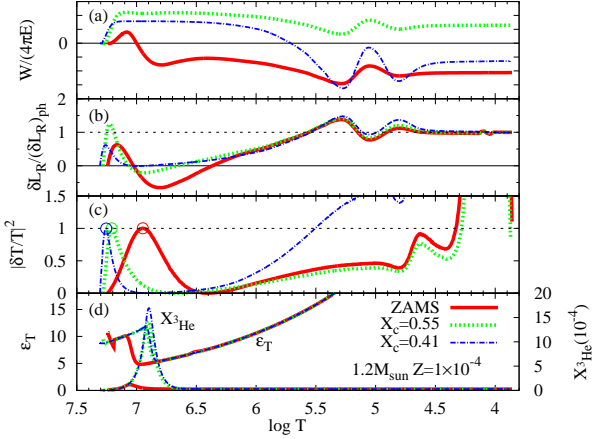}
  \end{center}
    \caption{Properties of models and the dipole g$_1$-mode at different evolutionary stages for the $1.2M_{\sun}$ star. {\bf (a)} work integral normalized with the total oscillation energy in a whole star. {\bf (b)} radiative luminosity perturbation normalized with the value at the photoshere. {\bf (c)} squared temperature perturbation normalized with the peak value in the deep interior, marked with the open circle. {\bf (d)} effective temperature dependence of nuclear reactions, $\varepsilon_T$, and $^3$He mass fraction ({\it A color version of this figure is available in the online journal}).}
    \label{fig:5}
\end{figure}

Table \ref{tab:1} and Fig. \ref{fig:5} show properties of the g$_1$-mode for the 1.2$M_{\sun}$ star with $Z=1\times 10^{-4}$ at different evolutionary stages. The core of the metal-poor low-mass stars is convective at the ZAMS stage even with the pp-chain burning because of the high central temperature, and the gravity waves are evanescent there. 
Note that the convective core might hardly affect the vibrational stability since convective time-scale there ($\sim$ yr) is much longer than the oscillation period ($\sim$ hr). In such a situation, it is plausible that convective flux does not react to pulsation \citep[e.g.][]{Goldreich1977, Pesnell1987, Guzik2000}.
As the convective core shrinks with stellar evolution, the gravity waves become to propagate in the deep interior. The g mode then becomes to have a large amplitude around the $^3$He shell, and unstable due to the $\varepsilon$-mechanism of the $^3$He-$^3$He reaction. 
For the higher mass stars, on the other hand, the pp-II and III branches and the CNO-cycle cotribute to the nuclear energy generation due to high temperature. Collisional reactions belonging to them have high temperature dependence comparable with that of the $^3$He-$^3$He reaction, and contribute to the g mode instability. 
In the later stage, however, the g mode becomes to have a p mode-like behaviour in the envelope. Then, flux dissipation there exceeds the excitation by the $\varepsilon$-mechanism, and the g mode becomes stable again.

\begin{figure}
  \begin{center}
    \includegraphics[width=84mm]{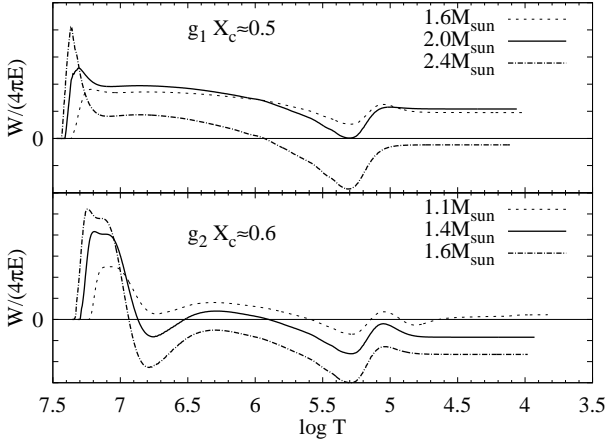}
  \end{center}
    \caption{Work integrals for the g$_1$ and g$_2$-modes for different mass stars with $Z=1\times 10^{-4}$.}
    \label{fig:6}
\end{figure}

We found that the above instability appears for stars, for example, with $\la 2.3M_{\sun}$ in the $Z=1\times 10^{-4}$ case. The more massive stars, on the other hand, keep substantial size of the convective core because of the dominant contribution of the CNO-cycle rather than the pp-chain. In this situation, the gravity waves cannot propagate in the nuclear burning core enough for the vibrational instability. Instead, just outside of the nuclear burning core, the amplitude of the g mode is relatively large, and strong damping is induced as shown in figure \ref{fig:6}. The g modes are then not destabilized indeed in those stars.    
  
\begin{table}
\caption{Properties of the g$_1$-mode shown in figure \ref{fig:5}. \label{tab:1}}
\begin{center}
\begin{tabular}{ccccccc} \hline\hline
  $X_c$      & age              &period& growth time scale \\ 
             & (yr)             &(hr)  &  (yr)             \\ \hline
  0.75       & $1.1\times 10^7$ & 1.78 & $-1.9\times 10^7$ \\
  0.55       & $8.0\times 10^8$ & 1.33 & $2.6\times 10^7$  \\
  0.41       & $1.2\times 10^9$ &0.997 & $-1.6\times 10^7$ \\ \hline
\end{tabular}
\end{center}
\end{table}

For the less massive stars, for example, ones with $\leq 1.1M_{\sun}$ in the g$_1$-mode case, the growth rate strongly increases in the late stage of the core hydrogen burning (figure \ref{fig:4}). Since such stars are located inside the classical instability strip, the g mode is excited by the $\kappa$-mechanism of helium ionization. As the g mode becomes to have a p mode-like behaviour in the envelope at the later stage, the $\kappa$-mechanism becomes to work efficiently, and leads to the stronger instability.  

Although the g$_2$-mode is also excited by the $\varepsilon$-mechanism, its instability appears mainly for the lower mass stars. Since the damping just outside the nuclear burning core is stronger compared with the g$_1$-mode, the g$_2$-mode is not destabilized for the higher mass stars (figure \ref{fig:6}). For the lower mass stars, on the other hand, such damping is weaker, and the g$_2$-mode is excited mainly by the $\varepsilon$-mechanism.

The g$_2$-mode instability appears also in the late stage of $\simeq 1.6M_{\sun}$ stars with $Z=1\times 10^{-4}$. In this case, although the convective core size is not substantial, the contribution of the CNO-cycle becomes to exceed that of the pp-chain, and hence $\mu$-gradient around the nuclear burning core becomes steeper. In this situation, gravity waves are trapped in the steep $\mu$-gradient zone, and the $\varepsilon$-mechanism works efficiently.

\subsection{Metallicity dependence}
\label{sec:metallcity}

\begin{figure}
  \begin{center}
    \includegraphics[width=84mm]{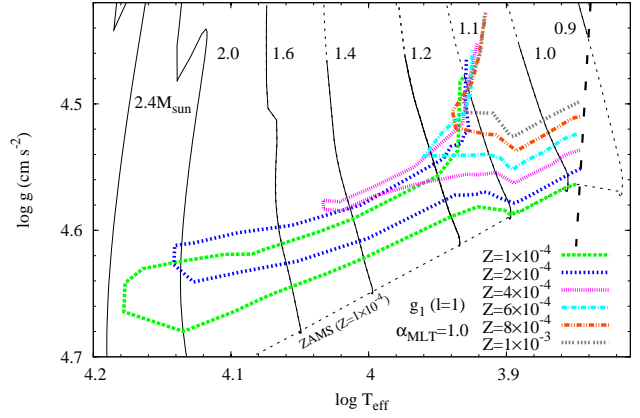}
    \includegraphics[width=84mm]{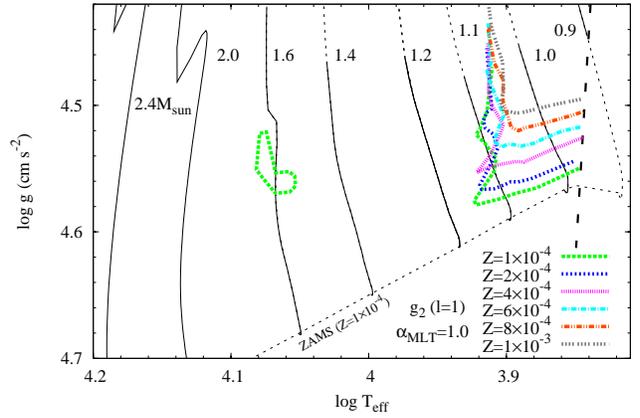}
  \end{center}
    \caption{Boudaries of instability regions for the dipole g$_1$ and g$_2$-modes on the $T_{\rm eff}$--$g$ plane with different metallicity. The evolutionary tracks are for $Z=1\times 10^{-4}$. The solid parts of the tracks correspond to stages at which the convective envelope is confined to the layer above $r/R=0.99$, while the dashed parts to those at which the convective envelope breaks in below $r/R=0.99$. The thick dashed line corresponds to the one in figure \ref{fig:2}. ({\it A color version of this figure is available in the online journal}).}
    \label{fig:7}
\end{figure}  

\begin{figure}
  \begin{center}
    \includegraphics[width=84mm]{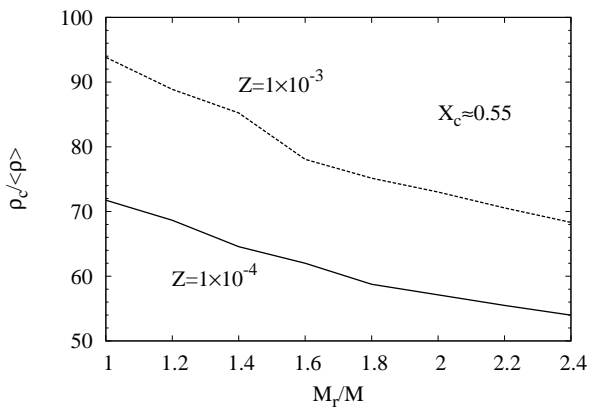}
  \end{center}
    \caption{Ratio of central density to average density, $\rho_c/\langle\rho\rangle$ for models with $Z=1\times10^{-3}$ and $1\times10^{-4}$ at an evolutionary stage $X_c\simeq 0.55$. The abscissa is stellar mass.}
    \label{fig:8}
\end{figure}

Figure \ref{fig:7} shows boudaries of instability regions on the $T_{\rm eff}$--$g$ plane for the g$_1$- and g$_2$-modes. They are cut off in the lower temperature side, where convective envelope becomes to extend below $r/R=0.99$. In the $Z=1\times 10^{-3}$ case, the instability appears for stars with $\la 1.2M_{\sun}$, and is induced mainly by the $\kappa$-mechanism. As the metallicity decreases, the instability region, particularly for the g$_1$-mode, extends toward the more massive stars. 

One of reasons for this is that metal-poorer stars have lower density contrast between the inner and outer regions than metal-rich stars. Figure \ref{fig:8} shows the ratio of the central density to the average density in a whole star, which represents the density contrast, for $Z=1\times 10^{-3}$ and $1\times 10^{-4}$. The ratio for metal-poorer stars is lower than for the metal-rich stars having the same mass, since metal-poorer stars are more compact and then have higher average density, while the central density is not substantially different among the metallicity. Due to this, the g mode amplitude is relatively larger in the deep interior, and the $\varepsilon$-mechanism can work efficiently for metal-poorer stars.    

Another reason is that as the metallicity decreases the CNO-cycle becomes replaced by the pp-chain for more massive stars. As discussed in Sec. \ref{sec:variation}, dominance of the pp-chain is favorable for the $\varepsilon$-mechanism instability, while CNO-cycle burning keeps substantial size of a convective core during stellar evolution.  

\begin{figure}
  \begin{center}
    \includegraphics[width=84mm]{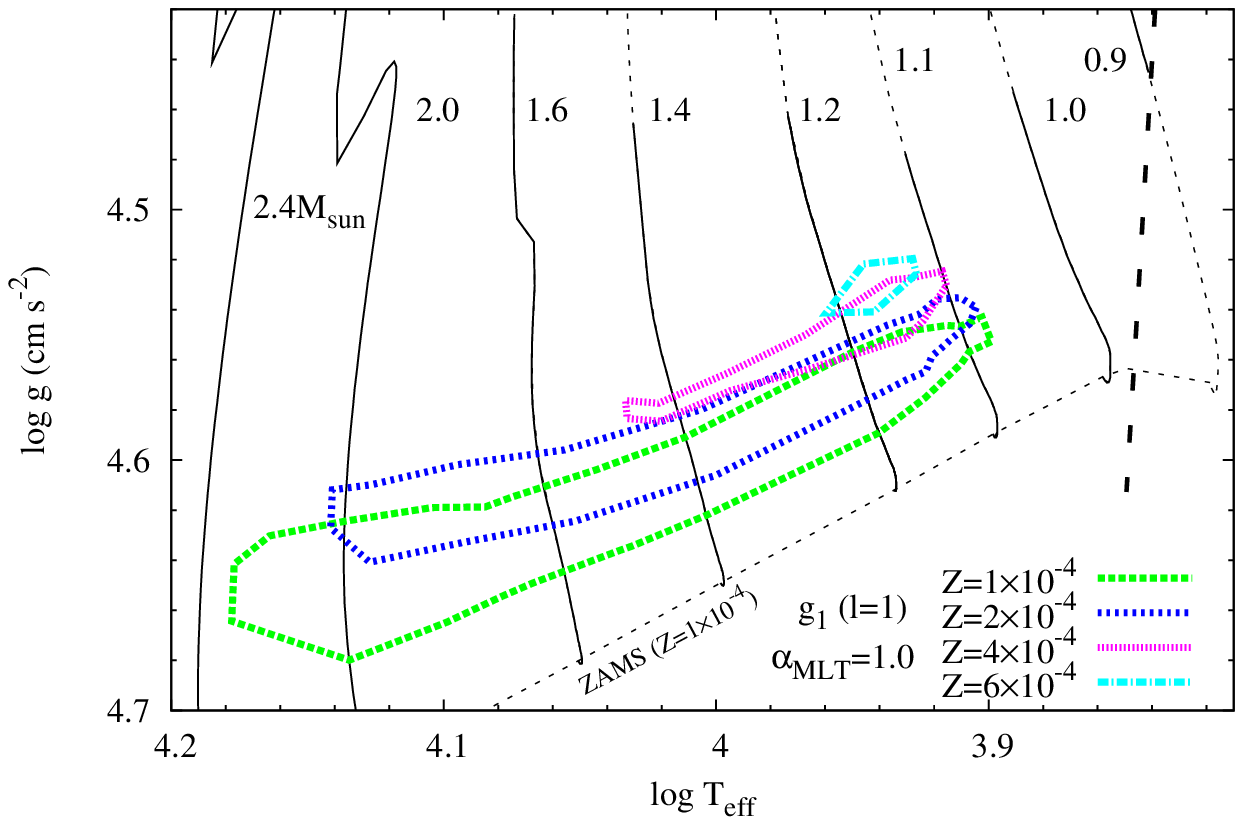}
    \includegraphics[width=84mm]{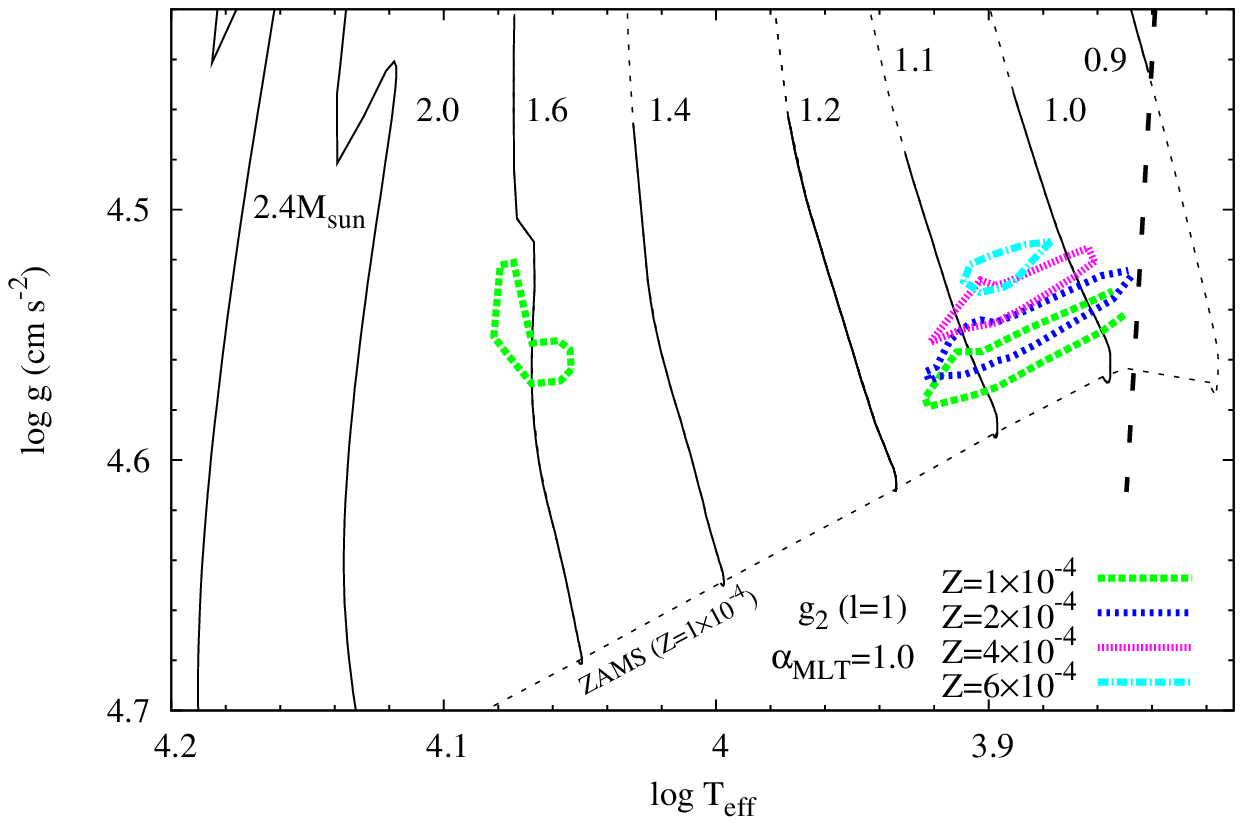}
  \end{center}
    \caption{Same as figure \ref{fig:7}, but for the instability with more than 50\% contribution of the $\varepsilon$-mechanism to the total excitation energy. ({\it A color version of this figure is available in the online journal}).}
    \label{fig:9}
\end{figure}

Figure \ref{fig:9} shows regions corresponding to the instability with more than 50\% contribution of the $\varepsilon$-mechanism to the total excitation energy. Such instability appears with $Z \la 6\times 10^{-4}$, and with $\log T_{\rm eff}\ga 3.90$ in the g$_1$-mode case. In this temperature range, corresponding to the bluer outside of the classical instability strip, the $\kappa$-mechanism of helium ionization does not work efficiently, and the $\varepsilon$-mechanism is responsible for the instability. In the g$_2$-mode case, on the other hand, such instability appears inside of the classical instability strip, while not in the bluer side except for the late stage of $\simeq 1.6M_{\sun}$ stars with $Z=1\times 10^{-4}$, at which the $\varepsilon$-mechanism instability is induced because of the steep $\mu$-gradient as described in Sec. \ref{sec:variation}. As shown in figure \ref{fig:6}, the g$_2$-mode is strongly excited by the $\varepsilon$-mechanism in the lower mass star case, while strong damping just outside of the nuclear burning core avoids the instability in the higher mass star case.

\section{Discussion}
Due to the $\varepsilon$-mechanism instability, the corresponding stars might exhibit pulsations with period $\sim$ 1 hr. The g modes have relatively substantial amplitude 
%of the luminosity perturbation 
at the surface as shown in figure \ref{fig:5}, although a nonlinear analysis is necessary to obtain the absolute value of the amplitude. 

\citet{Dziembowski1982, Dziembowski1983} estimated the dipole g$_1$-mode amplitude in the solar photosphere to be about 20 cm s$^{-1}$ by adopting the three mode coupling theory. In particular, \citet{Dziembowski1983} considered the special cases of parametric resonance, in which a parent linearly unstable mode is coupled with two dauther linearly stable modes. The oscillation energy of the parent mode is converted into that of the two daughter modes at a rate proportional to the product of the amplitudes of the three modes. The most effective for low-degree low-order parent g modes appears to be a coupling with a pair of similar high-degree daughter g modes to a state in which the amplitudes of all the three modes are steady.

The value 20 cm s$^{-1}$ is too small that the g mode oscillations are detectable for stars other than the Sun. On the other hand, SPB stars and $\gamma$ Doradus stars exhibit g mode oscillations with much higher amplitude. \citet{Appourchaux2010} suggested that such stars have only shallow convection zones, and therefore daughter g modes propagate much higher in the envelope and dissipate much more strongly, thus being limited themselves to much lower amplitudes and thereby being less able to extract energy from their parent. Such a situation also might be valid for the g modes in the metal-poor low-mass main-sequence stars analyzed in this study, although the linear growth rate is much smaller than in the SPB and $\gamma$ Doradus cases.
It is worth searching stellar pulsations induced by the $\varepsilon$-mechanism since such pulsations have not been detected yet and would be a great tool to examine the first derivative of the nuclear energy generation rate.

Metal-poor stars have attracted attention as clues to chemical evolution of the Universe. Thanks to advance in the observational side, a lot of metal-poor stars have been detected to date \citep[e.g.][]{Suda2008}. Although most of metal-poor stars are very far from the Sun, faint metal-poor pulsating stars have become detected in addition to the Cepheids and RR Lyr stars in the other galaxies, e.g., 20 $\delta$ Sct stars at $V\sim 23$ in the Carina dwarf spheroidal galaxy at $\sim 100$ kpc \citep{Mateo1998}. Recently, higher number of faint variables are found in Large Magellanic Cloud through the OGLE project \citep{Poleski2010}. 

Most of metal-poor stars are thought to be considerably old or to have the equivalent age with the Universe. The $\varepsilon$-mechanism instability reported in this paper, however, appears in the early stage of the main-sequence. In some galaxies such as dwarf Irregulars, on the other hand, the chemical evolution might be delayed and metal-poor star formation regions remain still now. In such galaxies, young metal-poor stars may exist and are expected to be candidates for the $\varepsilon$-mechanism pulsators.  

We also state a possibility that growth of the amplitude due to the vibrational instability might induce material mixing and have a significant influence on the later evolution of the star, as was expected concerning the solar neutrino problem by \citet{Dilke1972}. In particular, if the mixing occurred in the core, the surrounding cooler and hydrogen riched matter would be incorpolated into the core. This phenomenon might induce rejuvenation of the star and prolong the lifetime. To obtain a solution to this problem, we have to persue the nonlinear evolution.

\section{Conclusion}
We performed a linear nonadiabatic stability analysis of metal-poor low-mass main-sequence stars. We restricted our analysis to evolutionary models having negligibly thin convective envelopes, for which we can obtain a definite conclusion on the stability analysis without uncertainty in treatment of a convective envelope. We found that the dipole g$_1$- and g$_2$-modes certainly become unstable due to the $\varepsilon$-mechanism for $Z\la 6\times 10^{-4}$. The mass range of this instability extends toward higher mass with decreasing metallicity. 
One of reasons for this is that for metal-poorer stars density contrast between the inner and outer regions is lower, and that the g mode amplitude is relatively larger in the nuclear burning core. 
Another reason is that as the metallicity decreases the CNO-cycle burning becomes replaced by the pp-chain burning, which is favorable for the $\varepsilon$-mechanism instability.
As a result of the instability, stellar pulsation or material mixing is expected to occur. To confirm those, we require highly precise observations of faint metal-poor stars, and a nonlinear analysis of the oscillations.
 
\section*{Acknowledgments}
The authors are grateful to the anonymous referee for his/her useful comments for improving this paper. This study adopted the MESA code to construct the stellar evolutionary models. The authors wish to acknowledge the \citet{Paxton2011} publication and MESA website (http://www.mesa.sourceforge.net) for providing their very useful evolution code to public as an open source. This study has been financed by Global COE Program ``the Physical Sciences Frontier'', MEXT, Japan.

\end{document}